\documentclass[acmtog]{acmart}

\usepackage{xcolor}
\usepackage[linesnumbered,ruled,vlined]{algorithm2e}

\usepackage{acro}

\DeclareAcronym{lpq}{
  short=LPQ,
  long=Live Polling Quiz,
}
\DeclareAcronym{cs}{
  short=CS,
  long=Computer Science,
}
\DeclareAcronym{ug}{
  short=UG,
  long=Under Graduate,
}

\begin{document}

\title[The Impact of Live Polling Quizzes]{The Impact of Live Polling Quizzes on Student Engagement and Performance in Computer Science Lectures}


\author{Xingyu Zhao}
\orcid{0000-0002-3474-349X}
\affiliation{%
  \institution{Department of Computer Science, University of Liverpool}
  \streetaddress{}
  \city{Liverpool}
  \state{United Kingdom}
  \postcode{L69 3BX}
}

\affiliation{%
	\institution{WMG, University of Warwick}
	\streetaddress{}
	\city{Conventry}
	\state{United Kingdom}
	\postcode{CV4 7AL}
}
\email{xingyu.zhao@warwick.ac.uk}



%

\begin{abstract}
Prior to the COVID-19 pandemic, the adoption of live polling and real-time feedback tools gained traction in higher education to enhance student engagement and learning outcomes. Integrating live polling activities has been shown to boost attention, participation, and understanding of course materials. However, recent changes in learning behaviours due to the pandemic necessitate a reevaluation of these active learning technologies.

In this context, our study focuses on the Computer Science (CS) domain, investigating the impact of Live Polling Quizzes (LPQs) in undergraduate CS lectures. These quizzes comprise fact-based, formally defined questions with clear answers, aiming to enhance engagement, learning outcomes, and overall perceptions of the course module. A survey was conducted among 70 undergraduate CS students, attending CS modules with and without LPQs. The results revealed that while LPQs contributed to higher attendance, other factors likely influenced attendance rates more significantly. LPQs were generally viewed positively, aiding comprehension and maintaining student attention and motivation. However, careful management of LPQ frequency is crucial to prevent overuse for some students and potential reduced motivation. Clear instructions for using the polling software were also highlighted as essential.

\end{abstract}





\maketitle


\section{Introduction}

Before the COVID-19, live polling and real-time feedback tools were popular in higher education classrooms as a way to boost student engagement and performance in learning \cite{lim2017improving,serrano2019technology,voelkel2014new,lantz2014effectiveness,desouza2003comparison,salas2012analysis,hennig2019quizzing}. Research has shown that integrating live polling activities, where students respond to questions posed by the lecturers using digital devices, can increase student attention, participation, and understanding of the course materials \cite{bode2009clicker,kay2009examining,miller2013student}. Researchers suggest that by giving every student a chance to respond anonymously, live polling promotes inclusion and provides quieter students a voice \cite{kay2009examining}. The immediate feedback provided through live polling also allows lecturers to gauge student understanding in real-time and adjust their teaching timely \cite{voelkel2014new}. Live polling systems also create a more interactive learning experience that empowers students and keeps them actively involved during lectures \cite{lantz2014effectiveness}.

However, state-of-the-art research works on the impact of using live polling systems in in-person lectures were mostly performed based on data collected \textit{before} the COVID-19. Arguably, students' learning behaviours may have changed after the pandemic, following nearly 3 years without in-person teaching \cite{rapanta2021balancing}. The shift to remote learning during COVID-19 may have already altered student engagement and expectations \cite{cicha2021covid}. For instance, changes and new trends regarding the learning environment are identified in \cite{al2020future}, the new engagement behaviours in hybrid-classrooms is discussed in \cite{9815908}, and technology used in blended learning will likely play a critical role \cite{IMRAN2023100805}. In this regard, there is a need to re-examine student perceptions of active learning technologies, like live polling and real-time feedback tools, in light of these changes \cite{phelps2022using,reimers2021search}, which motivates this work\footnote{The work relates to A1, A3, V2, K3, K4 in UK-PSF \cite{advance_he_uk_2011} and Active Learning, Authentic Assessment and Digital Fluency in Liverpool Curriculum Framework \cite{the_centre_for_innovation_in_education_liverpool_nodate}.}.

Focusing on the Computer Science (CS) domain, we have undertaken a survey-based investigation to reevaluate the effects of integrating Live Polling Quizzes (LPQs) into undergraduate (UG) CS lectures, where the questions are all based on factual information and are formally defined, having clear answers, as opposed to open-ended or opinion-based questions that lack definitive answers. Our study revolves around several research questions that delve into aspects such as student engagement, learning outcomes, ideal frequency and usability, as well as potential correlations with overall perceptions of the course module. More specifically, a survey of 14 questions (12 multiple choices with Likert scale answers and 2 fill-in-the-blank questions) was distributed to 70 UG students who have been attending UG CS modules with and without using LPQs. 30 responses (in which 28 are valid) were collected and analysed both quantitatively and qualitatively. Finally, the threats to validity are discussed.

The new data shows some general insights as follows:
\begin{itemize}
    \item Although LPQs may have contributed to higher attendance, they alone do not completely account for the higher lecture attendance rates. Other relevant factors may have contributed more than LPQs.
    \item LPQs were broadly viewed as beneficial for comprehending the lecture content, effective at keeping most students’ attention, motivation and participation.
    \item However, increased LPQs frequency may benefit some students, it risks overuse for others if not managed carefully and diminishes their motivation to study outside of class.
    \item Moreover, a clear instruction on using the polling software appears to be very important to maximise the positive effect of LPQs.
\end{itemize}

Next, we present the main method in Section \ref{sec_method}, including the aims, design context, description on sample, and an overview of analysis approaches. Then the survey results with discussions are presented in Section \ref{sec_results}. Threats to validity is discussed in Section \ref{sec_threats} and finally we conclude in Section \ref{sec_conclusion}.



\section{Method}
\label{sec_method}

This section provides details of the method used to capture the impact of using LPQs on student engagement and performance. 

\subsection{Aim}

The aim of this study is to assess student perceptions and self-reported behaviours related to the use of LPQs during CS lectures. Specifically, the study seeks to understand if the inclusion of LPQs increases student motivation, engagement, and sense of understanding and connection in the classroom. Additionally, we aim to examine if attendance and participation in lectures with LPQs is associated with students reporting more positive benefits compared to traditional lectures without these interactive features.

\subsection{Context}

The University of Liverpool is a public research university located in Liverpool, England. It is a member of the Russell Group of leading research-intensive universities in the United Kingdom. Students in their third year of an UG CS programme at the University of Liverpool are typically working towards a Bachelor of Science (BSc) degree in CS. The programme provides students with core computer science knowledge and skills such as programming, algorithms, data structures, databases, operating systems, and more. In their third year, students take advanced courses like formal methods, machine learning, and often work on a final year project where they apply what they've learned to develop a computing system or conduct research. The programme prepares students for careers in software engineering, data science, and other technology fields upon graduation.

\subsection{Design}

\begin{figure*}[h]
\centering
\includegraphics[width=0.8\linewidth]{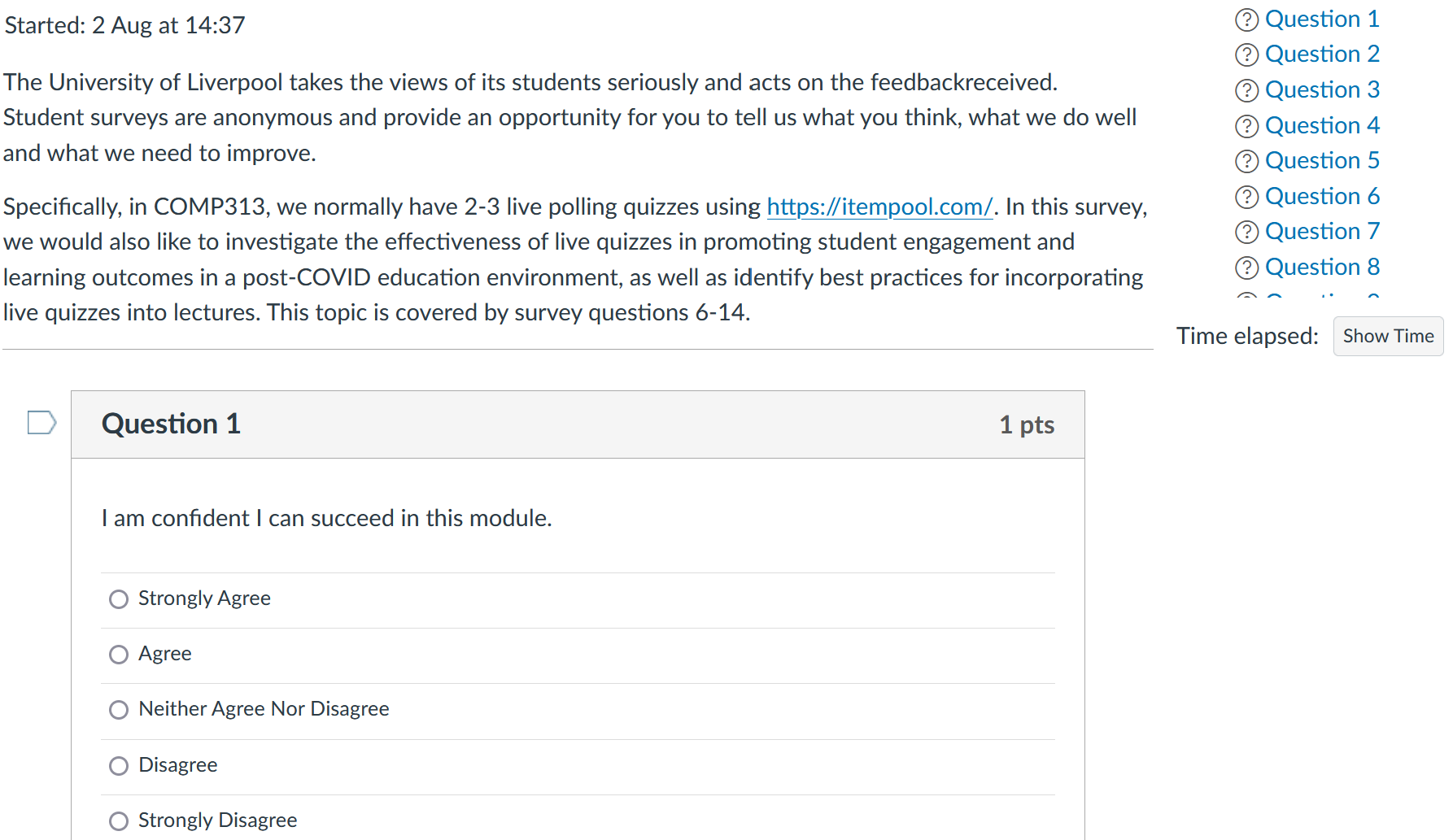}
\caption{UI of the survey in the CANVAS system}
\label{fig_ui_survey}
\end{figure*}

A survey with 14 questions (12 multiple-choice questions and 2 fill-in-the-blank questions), about 10 minutes in length, was designed to be completed online in the CANVAS system\footnote{A web-based teaching and learning management system used by the University.}. While the UI design of the survey is shown in Figure \ref{fig_ui_survey}, it was distributed, via links in emails and notifications from CANVAS, to the 70 CS UG students. The student may answer those questions anonymously and without any time limit in a three months time window after the mid-term.

\subsection{Sample}

\begin{figure}[h]
\centering
\includegraphics[width=1\linewidth]{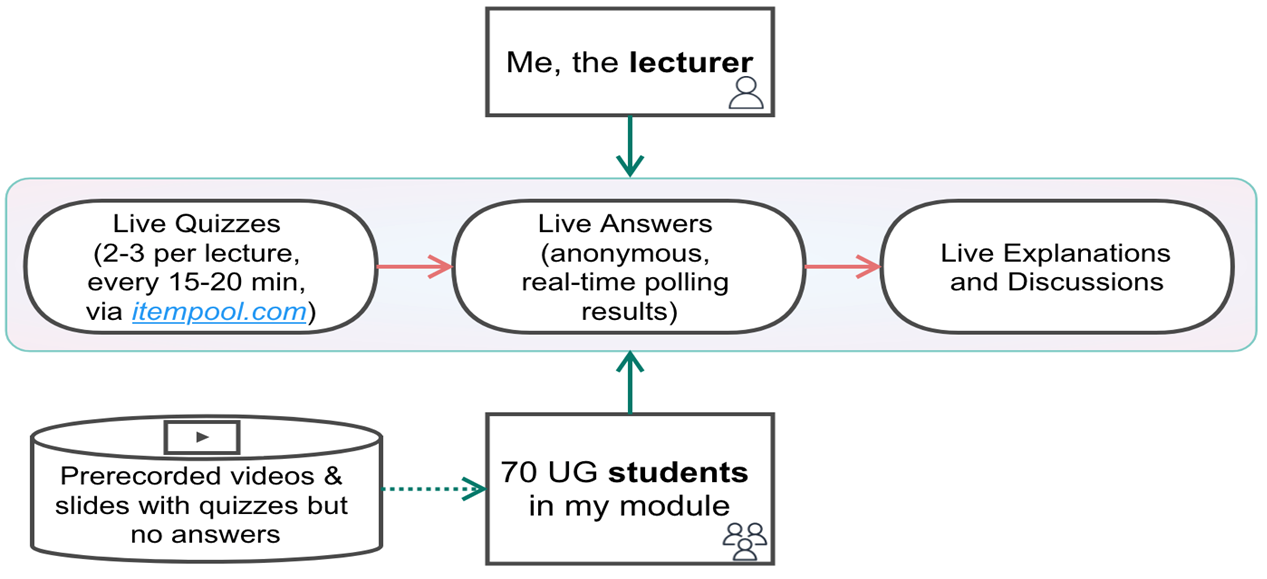}
\caption{The teaching practice around LPQs in the CS module where students were surveyed}
\label{fig_the_teaching_process}
\end{figure}

The survey was distributed to a cohort of 70 final year CS UG students in my own module, in which there are 3 lectures per week for 12 weeks. The general teaching practice around LPQs is shown in Figure \ref{fig_the_teaching_process}. Before each lecture, pre-recorded videos and slides with the quizzes (without answers) were sent to the students for pre-study. During the lectures, 2-3 LPQs were conducted every 15-20 minutes followed by live explanations and discussions of the answers. 

After the mid term when the students have sufficient experience of learning with my teaching practice with LPQs, the survey was distributed. Out of the 70 invited surveys, 30 responses were collected but with 2 invalid ones without any answers. That said, the later analysis is based on a sample of 28 valid responses, which was
deemed statistically adequate.

\subsection{Overview of analysis approaches}

For the 12 multiple-choice questions, we perform the following quantitative analysis:
\begin{itemize}
    \item Summarise and presenting the answer distribution of each question, showing the central tendency and spread of responses.
    \item Investigate correlations between questions using a correlation matrix of Pearson correlation coefficient \cite{cohen2009pearson}, and understand which questions have strong positive or negative correlations.
\end{itemize}

For the 2 fill-in-the-blank questions, not all respondents provide informative answers. Thus, a simple qualitative analysis are conducted and reported in the next section.

\section{Survey Results and Discussions}
\label{sec_results}

While all survey data and quantitative analysis code are publicly available at our project repository\footnote{ \url{https://github.com/x-y-zhao/LPQ\_survey\_study}}, we present and discuss the quantitative and qualitative analysis results in this section. 

\subsection{Data with basic statistics}

For the 12 multiple-choice questions, the first 4 questions (Q1-Q4) are about general feedback regarding the module, while the last 8 questions (Q5-Q12) are more specifically designed to assess perspectives on LPQs. This division of question types may later reveal correlations between students' overall perceptions of the module and their specific reflections on the use of LPQs.

\begin{figure}[h!]
\centering
\includegraphics[width=1\linewidth]{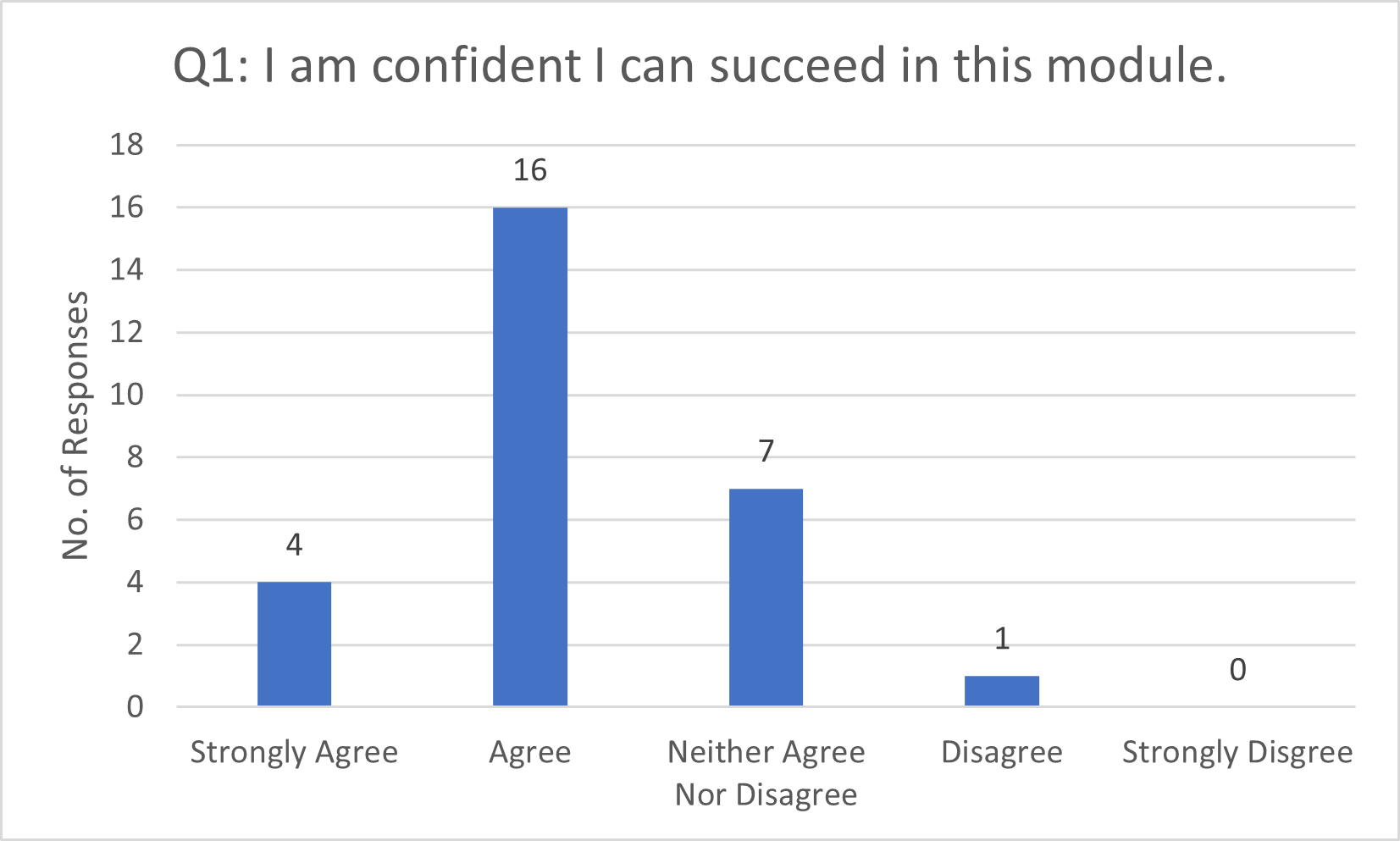}
\caption{Survey result of Q1.}
\label{fig_q1}
\end{figure}

\paragraph{Q1: I am confident I can succeed in this module.} The majority of respondents (about 71.4\%) expressed agreement or strong agreement with this statement, cf. Figure \ref{fig_q1}. This suggests most students feel confident in their ability to succeed in the module. However, about 28.6\% were neutral or expressed disagreement. This indicates a substantial minority may have lower self-efficacy. This distribution suggests that while most students are confident, we should be aware that some may need more support in developing self-efficacy. Targeted interventions could identify and assist students with lower confidence levels. More importantly, regarding LPQs, we will later exam if the level of confidence correlated with how student perceive the use of LPQs.

\begin{figure}[h!]
\centering
\includegraphics[width=1\linewidth]{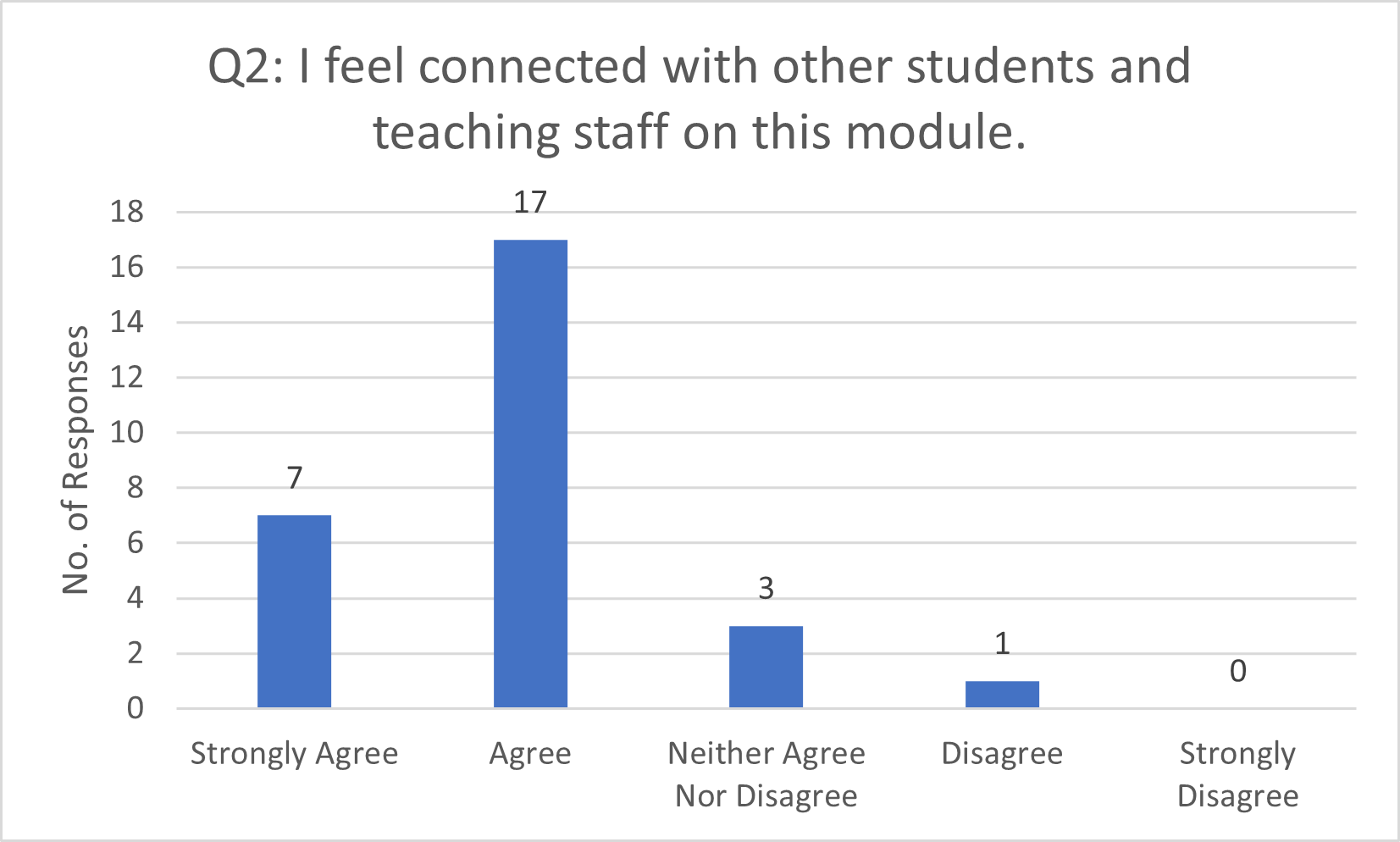}
\caption{Survey result of Q2.}
\label{fig_q2}
\end{figure}

\paragraph{Q2: I feel connected with other students and teaching staff on this module.} Responses were more positive on this question compared to Q1. Figure \ref{fig_q2} shows that around 85.7\% agreed or strongly agreed they felt connected. A small portion (10.8\%) were neutral on their feelings of connection. This suggests they did not perceive a strong sense of connection, but did not actively feel disconnected either. Only 1 respondent disagreed to some extent, signalling they felt disconnected from peers and the lecturer. Again, it would be very useful to later exam if the feeling of connected correlated with how student perceive the use of LPQs.

\begin{figure}[h!]
\centering
\includegraphics[width=1\linewidth]{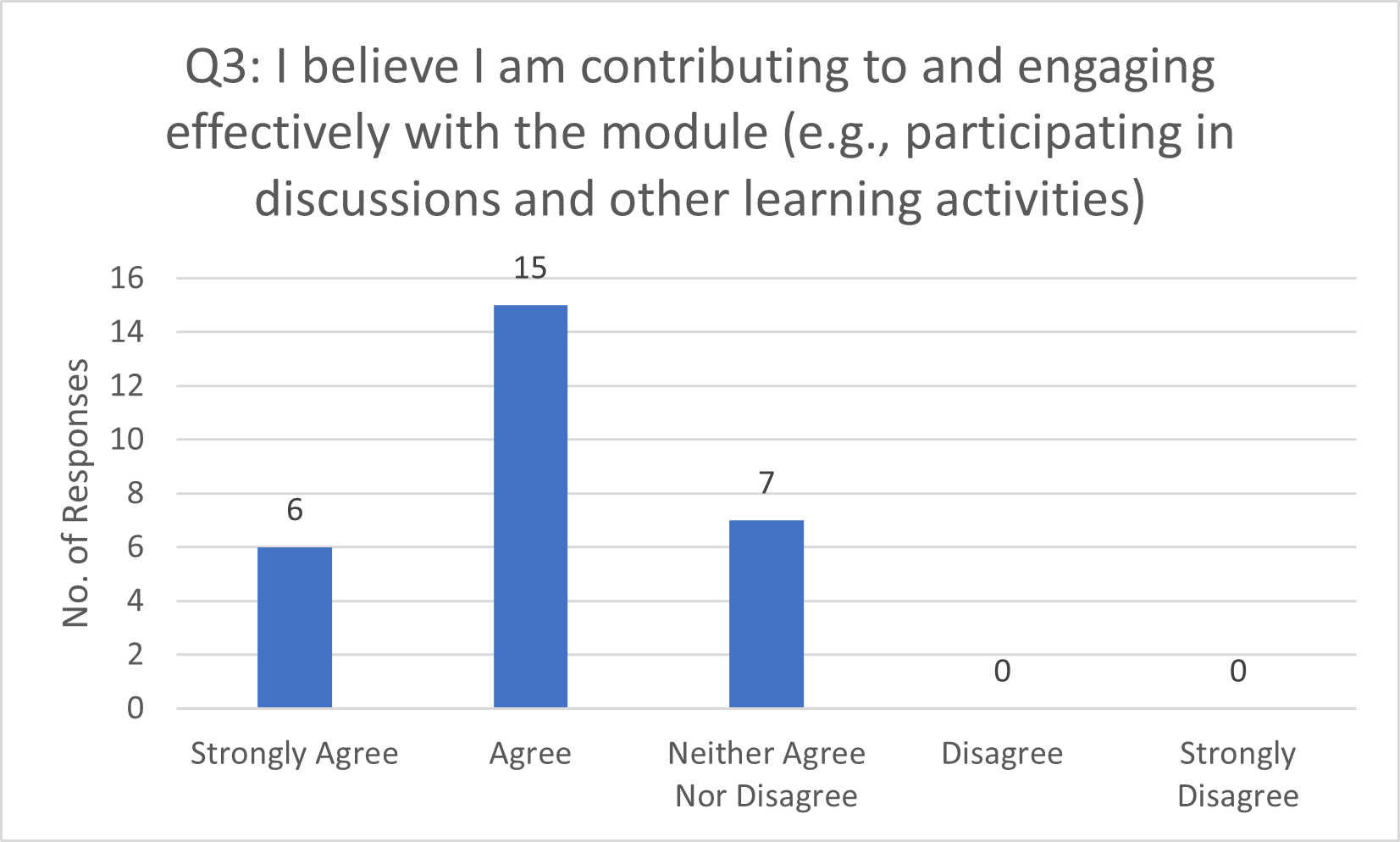}
\caption{Survey result of Q3.}
\label{fig_q3}
\end{figure}

\paragraph{Q3: I believe I am contributing to and engaging effectively with the module (e.g., participating in discussions and other learning activities)} As shown in Figure \ref{fig_q3}, a majority of respondents, 75\%, agreed or strongly agreed that they were actively contributing and engaging with the module. The remaining 25\% were neutral, neither agreeing nor disagreeing about their participation. So in fact, most students responding to the survey did perceive that they were actively involved and participating in the module learning activities. This might hint the positive outcome of using LPQs, which needs more investigation later.

\paragraph{Q4: At this stage in the module I understand how my learning will be assessed on this module} Almost every respondent (27 out of 28) agreed or strongly agreed that they understand how their learning will be assessed in the module. Only 1 respondent was neutral, neither agreeing nor disagreeing with the statement. This indicates that, despite the extensive use of LPQs in lectures, it did not confuse students about how the module would be assessed at the end of the term. As long as clear statements about assessment are provided in the module specification and reiterated verbally during lectures, the inclusion of LPQs throughout the course does not appear to obscure the end-of-term evaluation methods.

\begin{figure}[h!]
\centering
\includegraphics[width=1\linewidth]{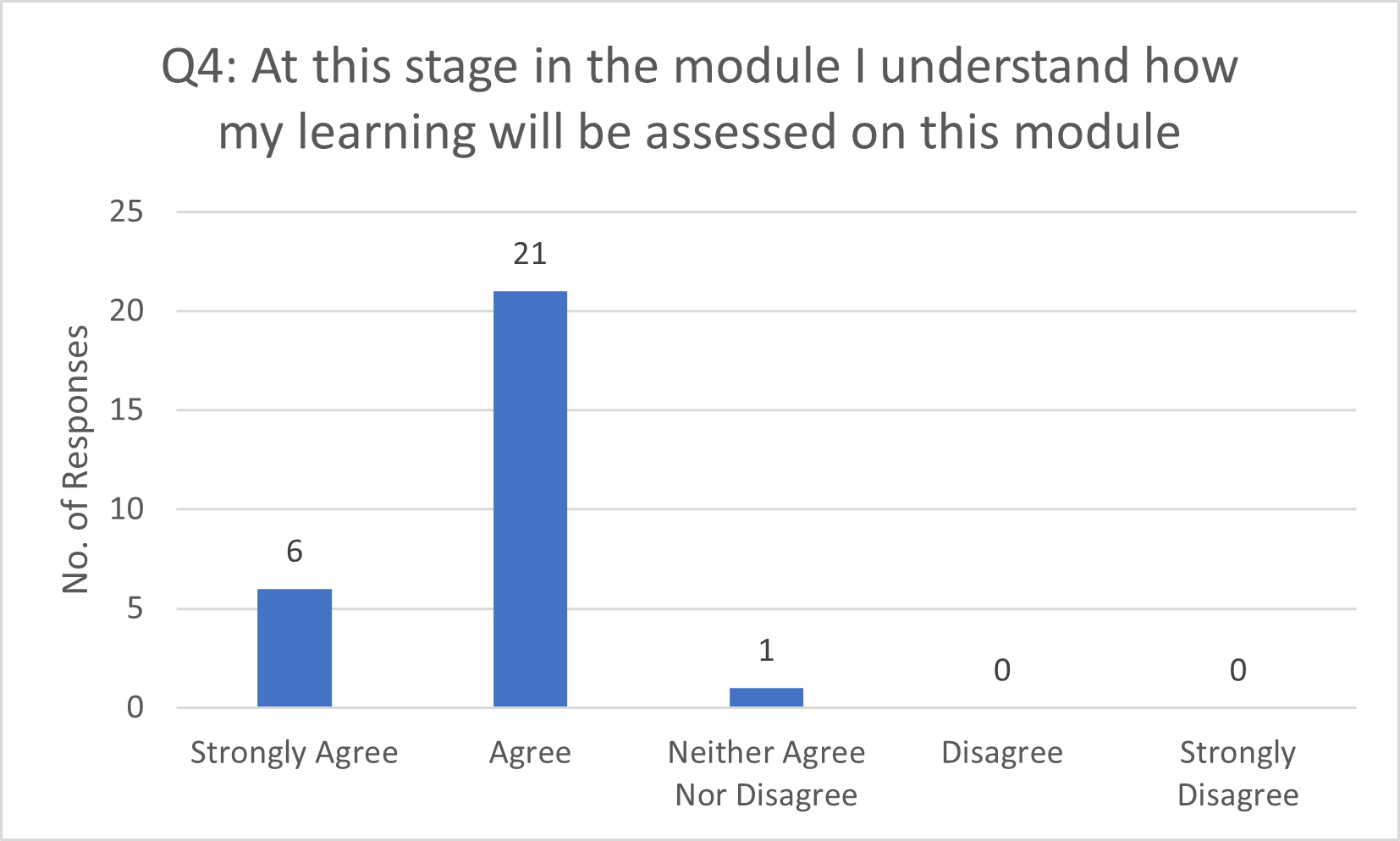}
\caption{Survey result of Q4.}
\label{fig_q4}
\end{figure}

\begin{figure}[h!]
\centering
\includegraphics[width=1\linewidth]{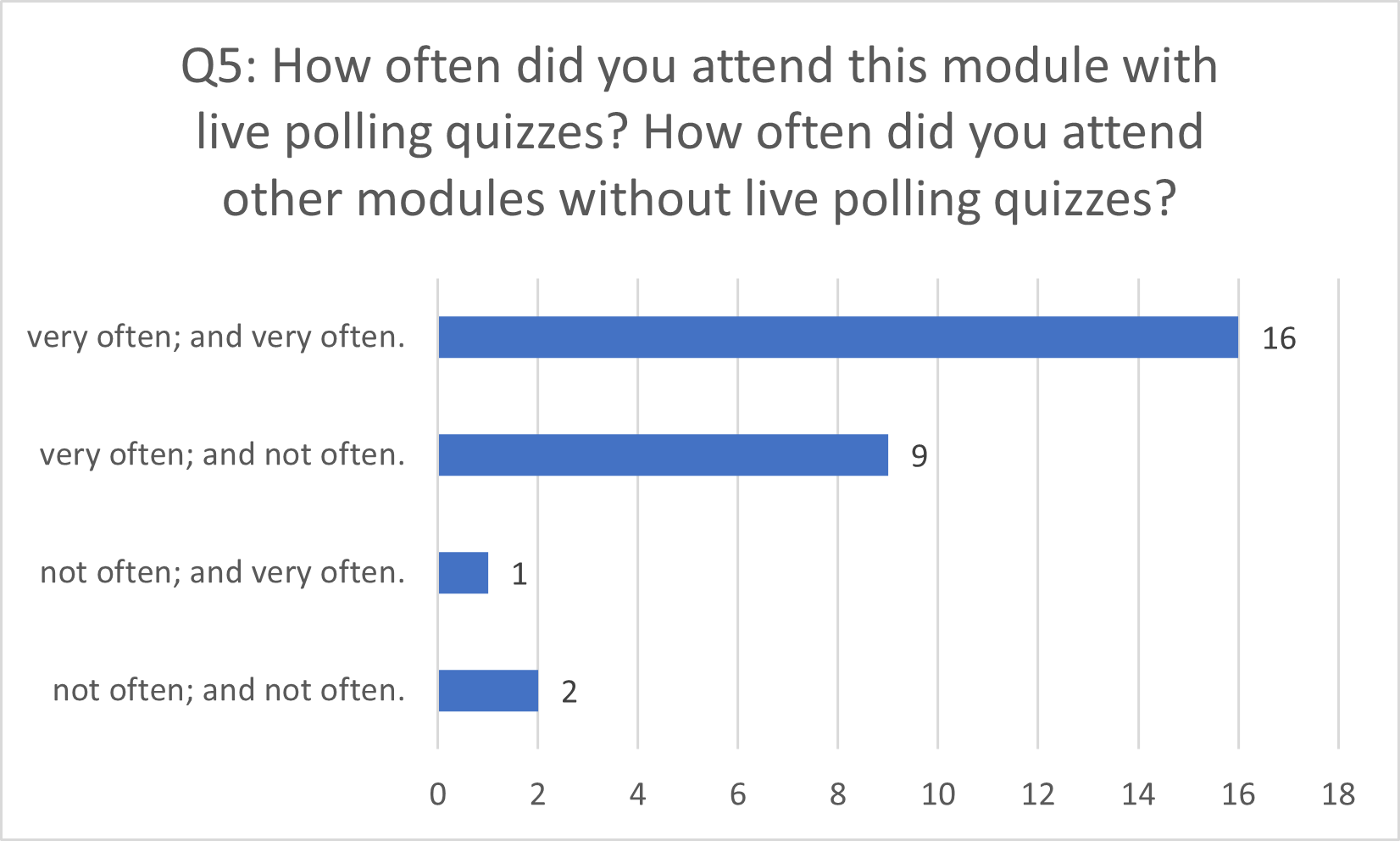}
\caption{Survey result of Q5.}
\label{fig_q5}
\end{figure}

\paragraph{Q5: How often did you attend this module with LPQs? And how often did you attend other modules without LPQs?} As shown in Figure \ref{fig_q5}, the majority (64.3\%) are neutral in terms of if the use pf LPQs helps the attendance. A substantial proportion (32.1\%) suggests the LPQs may have increased motivation and engagement, leading to higher lecture attendance rates. That said, 1 respondent attended other lectures more frequently without LPQs. So, while LPQs may have encouraged attendance, they do not fully explain the variation in lecture attendance rates. Other factors are likely also at play.

\paragraph{Q6: Did the LPQs help you better understand the lecture material?}  The vast majority of respondents (approximately 93\%) answered ``Yes'' that the LPQs helped them better understand the lecture material. Only a very small number answered ``Not sure", indicating they were uncertain about the impact. The overwhelmingly positive response suggests the LPQs were broadly viewed as beneficial for comprehending the lecture content. This aligns with the intent of using active learning techniques like polling \cite{draper2004increasing,arthurs2017integrative} to check understanding and clarify knowledge gaps during lectures.

\begin{figure}[h!]
\centering
\includegraphics[width=1\linewidth]{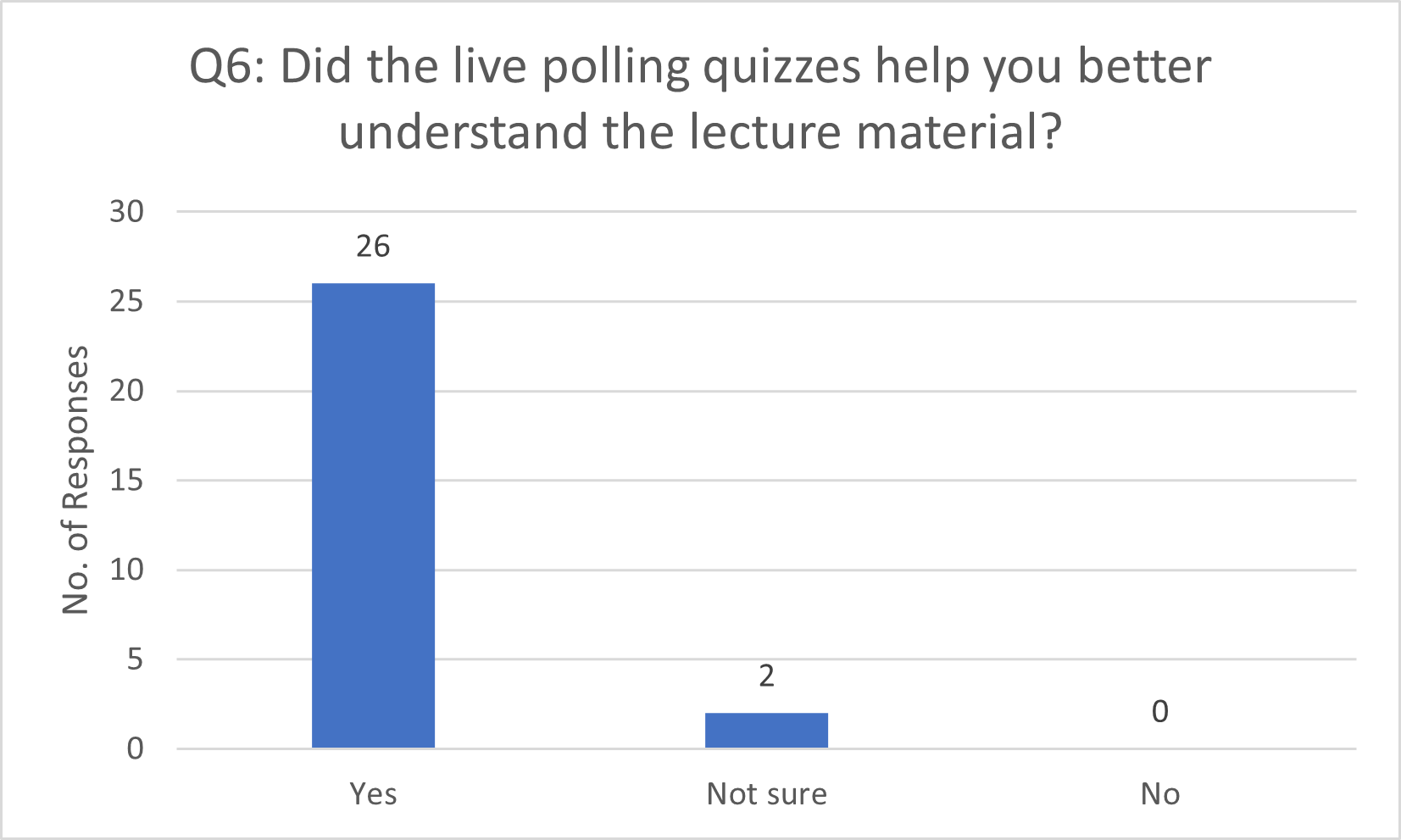}
\caption{Survey result of Q6.}
\label{fig_q6}
\end{figure}

\paragraph{Q7: Did the LPQs help you stay engaged during the lecture?} The majority of respondents (around 85.7\%) answered ``Yes'' that the LPQs helped them stay engaged during lectures.
However, the rest were uncertain. Overall, the responses indicate LPQs were broadly effective at keeping most students' attention and participation during lectures.

\begin{figure}[h!]
\centering
\includegraphics[width=1\linewidth]{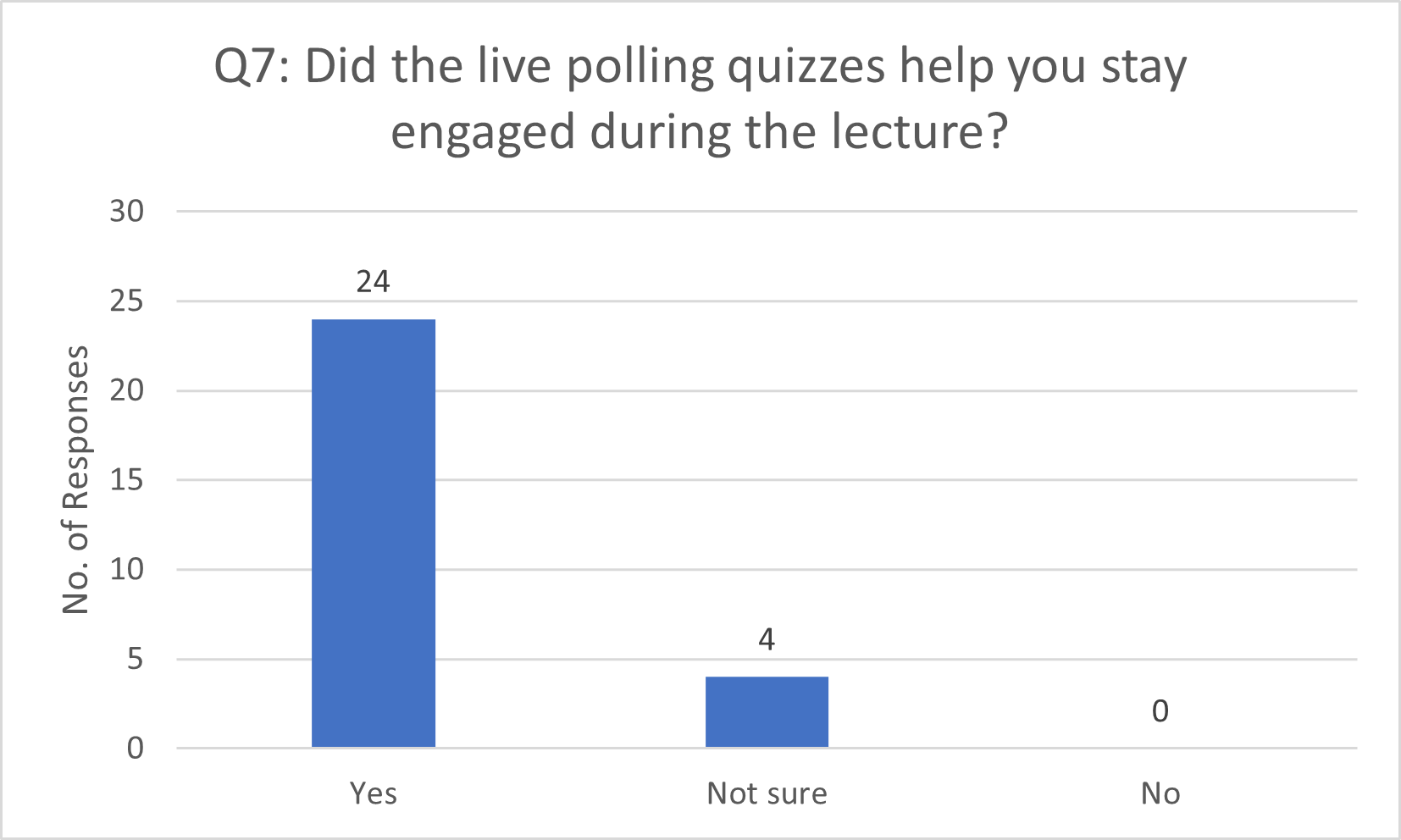}
\caption{Survey result of Q7.}
\label{fig_q7}
\end{figure}

\begin{figure}[h!]
\centering
\includegraphics[width=1\linewidth]{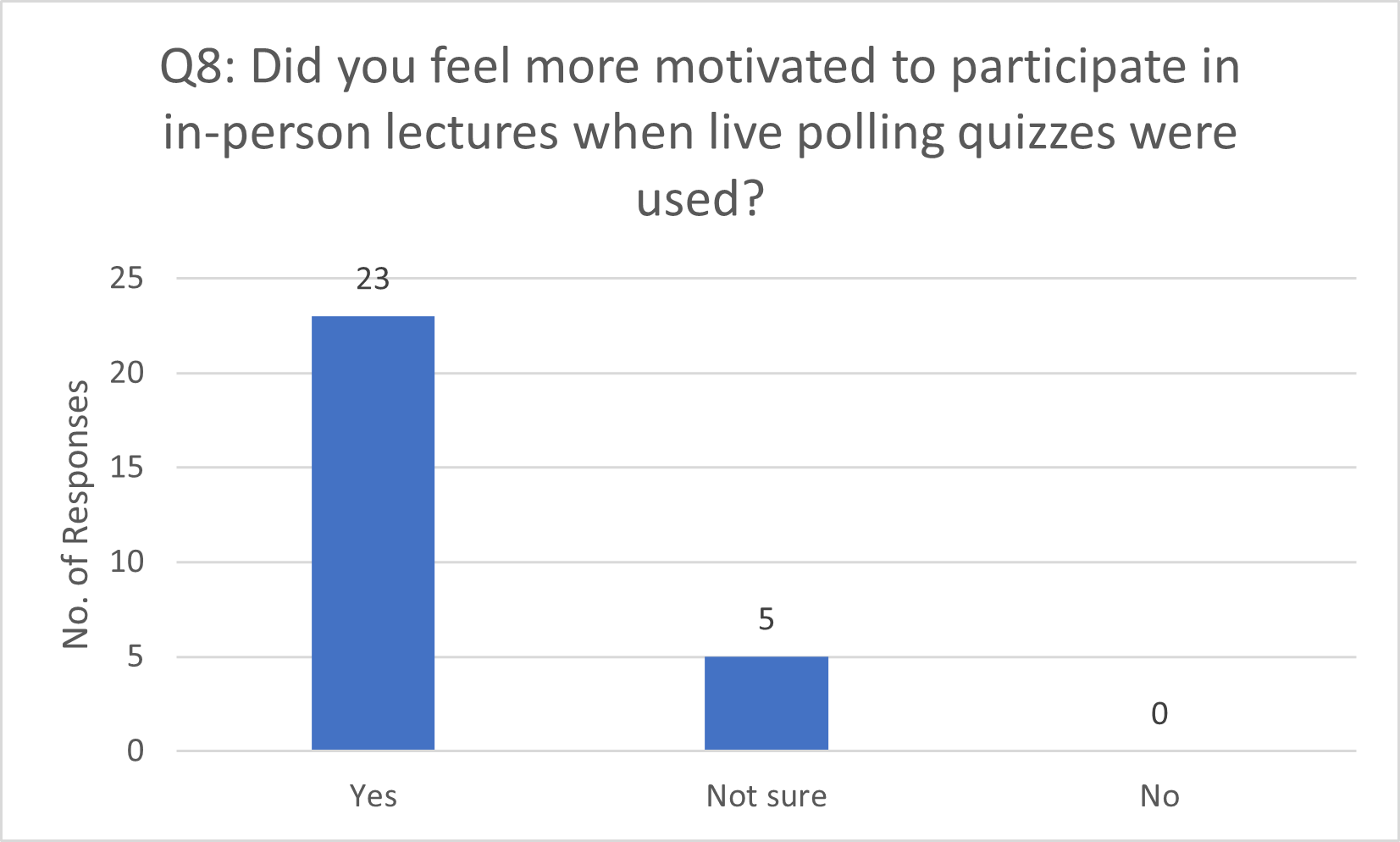}
\caption{Survey result of Q8.}
\label{fig_q8}
\end{figure}

\paragraph{Q8: Did you feel more motivated to participate in in-person lectures when LPQs were used?} 
Similarly to Q7, the use of LPQs motivated participation for most students, a few were uncertain on their motivation levels, as shown in Figure \ref{fig_q8}

\begin{figure}[h!]
\centering
\includegraphics[width=1\linewidth]{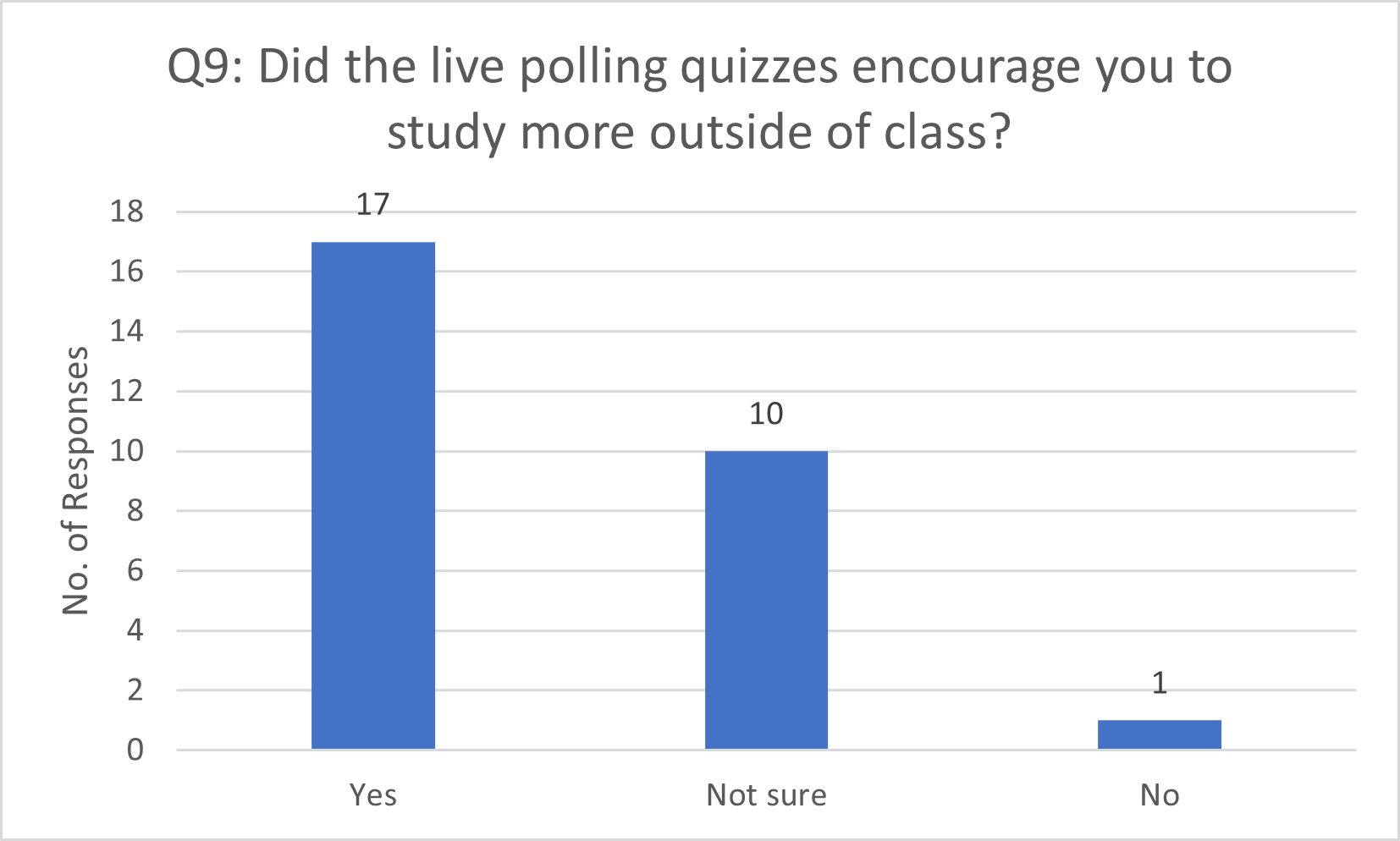}
\caption{Survey result of Q9.}
\label{fig_q9}
\end{figure}

\paragraph{Q9: Did the LPQs encourage you to study more outside of class?} Compared to the very positive feedbacks in Q7 and Q8, there is a clear drop in the number of responses with a positive ``Yes'', cf. Figure \ref{fig_q9}. Responses to this question were relatively mixed---60\% answered ``Yes'', indicating the quizzes encouraged them to study more outside of class, while 40\% were uncertain or feel LPQs did not affect their study habits outside classes. This indicates LPQs may have limited positive impact on their study habits outside classes.

\paragraph{Q10: Were the instructions for using the live polling software clear and easy to follow?} Analysing the responses, the majority (around 86\%) answered ``Yes'' that the instructions for using the polling software were clear and easy to follow.
However, around 24\% were ``Not sure''. This suggests that while the instructions seem adequately clear overall, there is room for improvement to make them more universally understandable. Moreover, we will exam how this correlates answers to other questions to understand the importance of a clear instruction of using the software.

\begin{figure}[h!]
\centering
\includegraphics[width=1\linewidth]{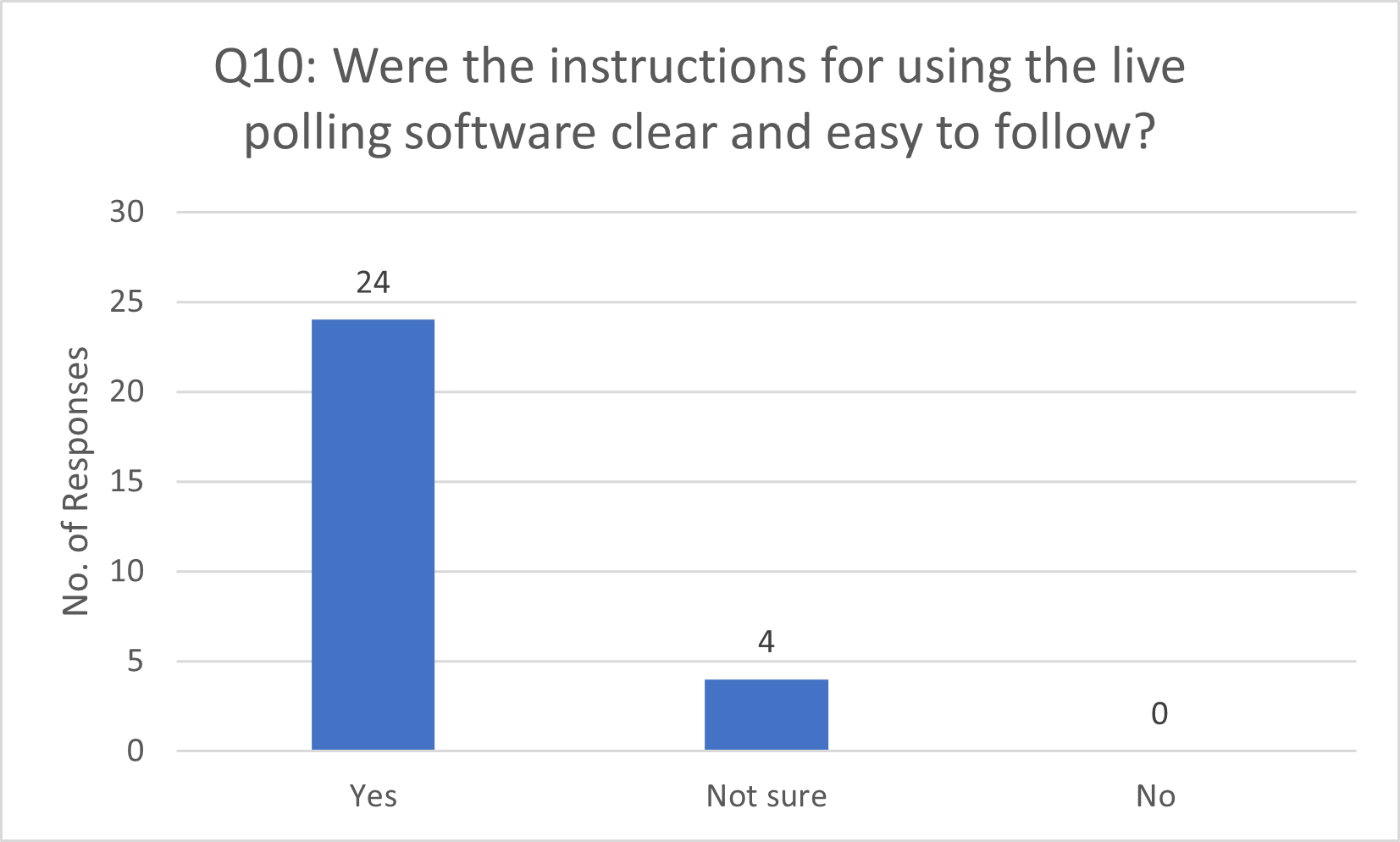}
\caption{Survey result of Q10.}
\label{fig_q10}
\end{figure}

\begin{figure}[h!]
\centering
\includegraphics[width=1\linewidth]{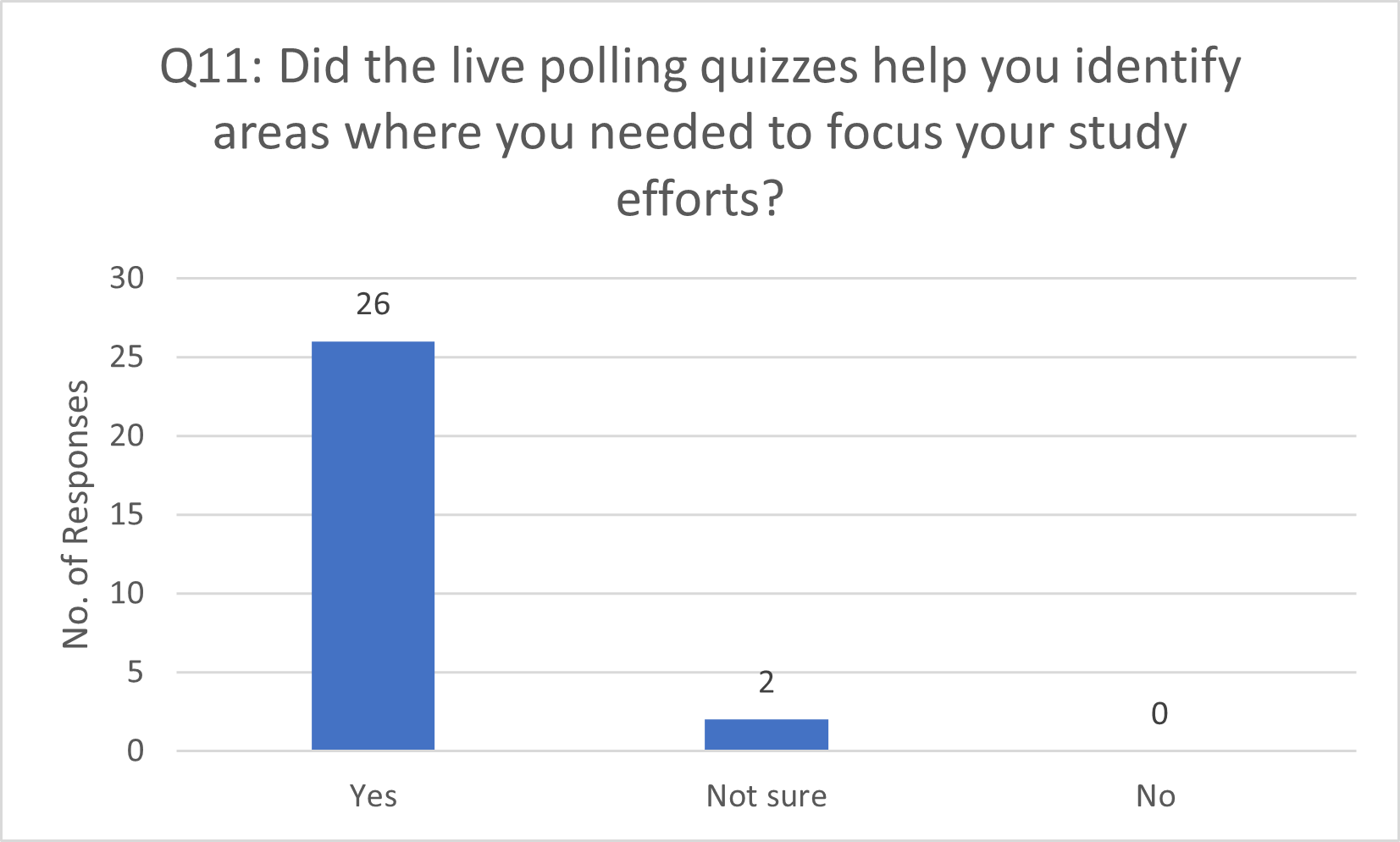}
\caption{Survey result of Q11.}
\label{fig_q11}
\end{figure}

\paragraph{Q11: Did the LPQs help you identify areas where you needed to focus your study efforts?} As presented in Figure \ref{fig_q11}, we got very positive outcome of this question---93\% respondents feel LPQs helped identify areas needing further study, while only 7\% are uncertain. We may confidently conclude that the majority of students found LPQs useful for highlighting areas to concentrate their studying on.

\begin{figure}[h!]
\centering
\includegraphics[width=1\linewidth]{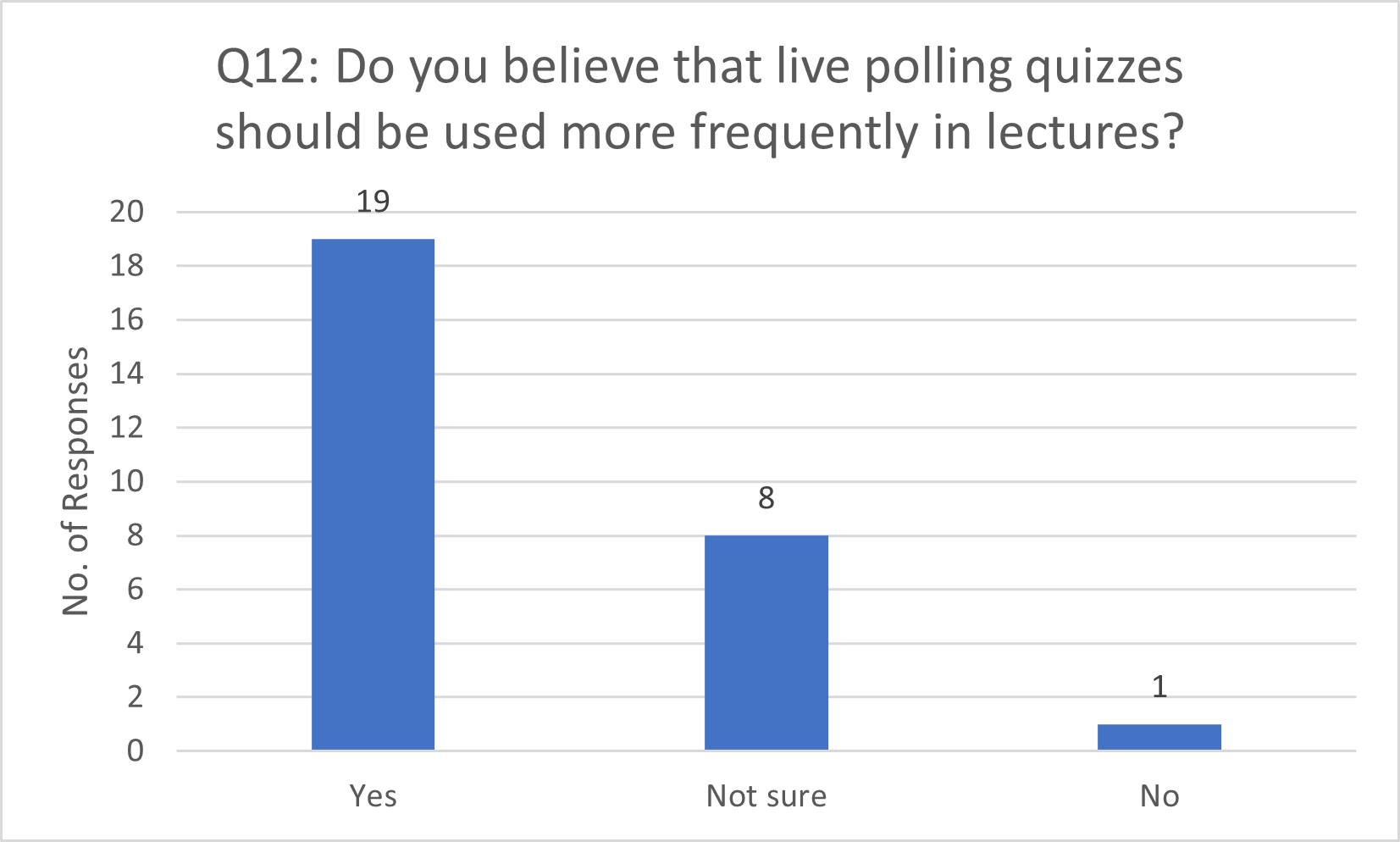}
\caption{Survey result of Q12.}
\label{fig_q12}
\end{figure}

\paragraph{Q12: Do you believe that LPQs should be used more frequently in lectures?} Interestingly shown in Figure \ref{fig_q12}, the responses were relatively mixed compared to previous questions, despite those positive aspects of using LPQs. Approximately 68\% of students answered ``Yes'' that LPQs should be used more frequently. Around 29\% were ``Not sure'' if increased usage was beneficial, and 1 answered ``No''. This suggests that increased LPQs frequency may benefit some students but risks overuse for others if not managed carefully.

\subsection{Correlation analysis}

To perform quantitative correlation analysis between the 12 multi-choice questions, we first encode the answers into numerical values:
\begin{itemize}
\item Strongly Agree = 1; Agree = 0.5; Neither Agree nor Disagree = 0; Disagree = $-0.5$; Strongly Disagree = $-1$
\item Very often and very often = 0; No often and not often = 0; Very often and not often = 1; Not often and very often = $-1$;
\item Yes = 1; Not Sure = 0; No = $-1$
\end{itemize}
Intuitively, we encode answers in favour of using LPQs as positive numbers, neutral answers as 0, and otherwise negative numbers.

\begin{figure}[h!]
\centering
\includegraphics[width=1\linewidth]{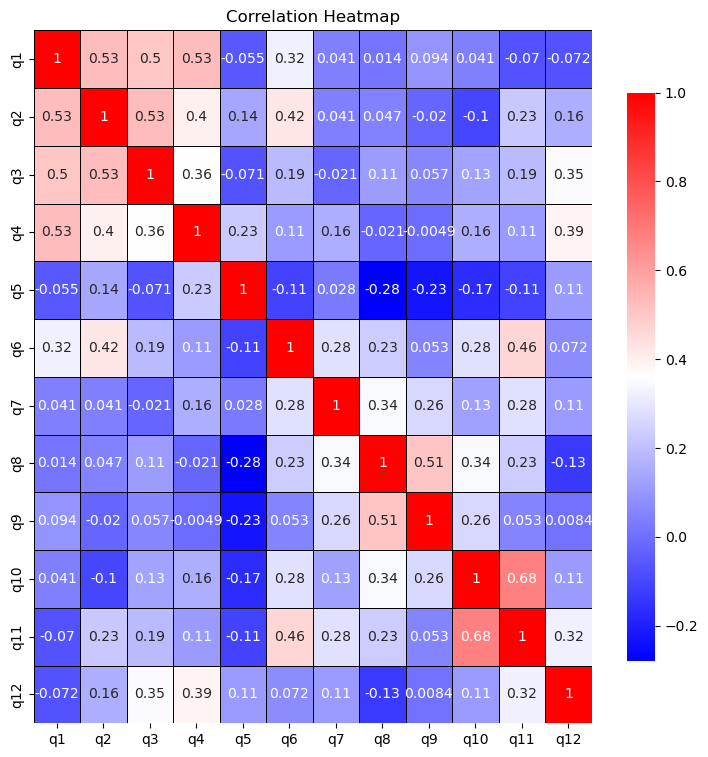}
\caption{The correlation matrix of the 12 multi-choice questions, showing the Pearson correlation coefficients.}
\label{fig_corr}
\end{figure}

Figure \ref{fig_corr} shows the correlation matrix, in which each correlation coefficient ranges from -1 to +1. A positive value indicates a positive correlation, a negative value indicates a negative correlation, and a value close to 0 indicates little to no linear correlation \cite{carlson2012understanding}. Although there are several correlation coefficients we can use, we present the most common Pearson correlation coefficient \cite{cohen2009pearson} in the matrix.

We first highlight those positively correlated questions/answers with sufficiently high degree ($>0.3$). As expected, the answers of the first 4 questions (Q1--Q4) regarding general feedbacks of the module are all strongly correlated, indicating the validity of the responses collected. The answers to Q1 and Q6, as well as Q2 and Q6, show a positive correlation. This correlation implies that the extent to which students use LPQs to understand learning material is positively associated with both their confidence in being successful and their sense of connection with others in the lectures. In other words, the more students effectively utilise LPQs, the more likely they are to feel confident about succeeding and connected with their peers during lectures.
Q7 and Q8 show a positive correlation, indicating that if LPQs help a student stay engaged during lectures, then they are also more likely to motivate that student to attend lectures in person. The positive correlations between Q8 and Q9, as well as between Q8 and Q10, indicate that feeling motivated by LPQs to attend lectures is positively associated with both being encouraged to study more outside of class and quickly adapting to using learning software. The positive correlation between Q11 and Q6 says when LPQs help the student better understand the learning material, it also identify the area where the student need to focus. Very significantly, Q11 and Q10 are positively correlated, showing the clear instruction on using software positively correlates LPQs help in identifying areas to focus. Finally, Q12 shows positive correlations with Q3, Q4, and Q11 respectively. This indicates that students who are in favour of using LPQs more frequently also tend to report higher engagement in the module, better understanding of how the module will be assessed, and that LPQs helped them identify areas to focus on. In other words, students who want more regular use of LPQs believe they lead to greater engagement, clarity on assessments, and ability to pinpoint knowledge gaps.

While generally all negative correlations among those 12 questions/answers are not very significant, we highlight two pairs that have correlations smaller than -0.2---Q5 and Q8, and Q5 and Q9. This indicates: i) Students who attend lectures with LPQs relatively more often than lecturers without LPQs do not necessarily feel more motivated because of the use of LPQs. ii) Students who attend lectures with LPQs relatively more often than lecturers without LPQs do not feel LPQs encourage them to study more outside of class. While a rigorous study with more data is needed to draw accurate conclusions, for now, we can interpret them as: i) LPQs might have encouraged attendance, but there are likely other relevant factors that contribute more than LPQs in this regard. ii) Students who attend in-person lectures with LPQs more often may rely too much on them, which diminishes their motivation to study outside of class.



\subsection{Qualitative analysis}

To prevent respondent fatigue and ensure data quality, limited number of fill-in-the-bank questions should be used in the survey \cite{galesic2009effects}. That said, we included the following two open questions.

\paragraph{Q13: What one thing can we do to better to improve your experience on this module?} This is a general question regarding how the students perceive the module. In-total 9 answers were given out of those 28 valid respondents, covering the points of asking for more exercises/tutorials, and more interesting examples.

\paragraph{Q14: What suggestions do you have for improving the use of LPQs in lectures?} This question is specifically designed for LPQs. In-total 10 answers were given out of those 28 valid respondents, and interestingly all the 9 respondents answered Q13 also answered Q14. The main points include: the link for the web-based LPQs should be more accessible, and more LPQs per lecture would be better.

Considering observations form previous sections, the qualitative results show that the use of LPQs in lectures has a potential influence on students' study behaviour and engagement. Implementing more easily accessible web-based LPQs with clear instructions of use (which is indeed a challenge as identified by \cite{kay2009examining}), and increasing the frequency of LPQs during lectures could contribute to a more engaging and effective learning environment for some students.

\section{Threats to Validity}
\label{sec_threats}

\paragraph{Construct Validity} It refers to the extent to which a survey instrument measures the theoretical construct or concept it claims to be measuring. In this study, the use of self-reported survey data is subjective and prone to self-reporting bias that may pose a threat. Students may have responded more positively due to wanting to please the lecturer who is also the researcher. To mitigate these two threats, we assure respondents that their responses will be kept confidential and anonymous. This can reduce the fear of social judgement and encourage more objective and accurate reporting.  Students who had more positive views on LPQs may have been more inclined to take the survey, while others with negative views may be underrepresented. To mitigate this response bias, we have sent reminder messages to non-respondents to encourage their participation during the 3-months survey period, and tried to clearly communicate the purpose of the survey and its importance to all students regardless of their views.

\paragraph{Internal Validity} Threats may correspond to bias in establishing cause-effect relationships in our survey. When interpreting the results of our correlation analysis, we are fully aware that ``correlation does not imply causation'' is a fundamental principle in research and statistics. Confounding variables, i.e., factors other than the use of LPQs (e.g., physical learning environment and teaching material quality) may have influenced outcomes like attendance and engagement. To mitigate such threat, we plan to do Randomised Controlled Trials in the future by carefully considering and addressing confounding variables in the design and analysis. In this preliminary study with limited data, we only draw correlation conclusion.

\paragraph{External Validity} Factors limiting generalisability threaten external validity. The selection bias (e.g., only a group of UG CS student from one University was study) and small sample size are two threats in this regard. In addition, only one polling software (itempool.com) was studied that also poses a threat.  To mitigate them, more sample needs to be collect with diversified data representing more students and software tools. In this preliminary study, we have explicitly discussed the limitations related to our sample's representatives.

\section{Conclusion}
\label{sec_conclusion}

 After nearly 3 years of remote education, student behaviours and preferences may have shifted. There is a need to re-evaluate live polling systems given changes brought on by COVID-19. This study aims to provide up-to-date insights into student perceptions of LPQs based on recent experience. Focusing on CS UG students, a survey was distributed to students who attended lectures both with and without LPQs. The quiz questions were factual and objectively answerable. The survey examined research questions around engagement, learning outcomes, ideal frequency/usability of polling, and correlations with overall course perceptions. 
 
 Preliminary findings indicate that while live polling contributed to lecture attendance and comprehension for most students, it did not wholly account for them. Increased frequency risks overuse and diminished motivation for some students. Clear software instructions were deemed very important. While students broadly saw polls as beneficial, COVID-era changes necessitate renewed investigation. 
 
This work addresses a literature gap by re-evaluating active learning technology amidst modern educational contexts shaped by the pandemic. Specifically it sheds light on the evolving role of LPQs in the post-COVID-19 CS education for UG students, providing valuable insights for educators and institutions seeking to optimise student engagement and learning experiences. 

Moving forward, we plan to work on ongoing monitoring and data collection to further mitigate the identified validity threats, along with conducting advanced quantitative analyses like clustering and factor analysis as more data becomes available.



\bibliographystyle{ACM-Reference-Format}
\bibliography{ref}



\end{document}